\documentclass[aps, prl, twocolumn, superscriptaddress]{revtex4}

\usepackage{amsfonts}           
\usepackage{amssymb}            
\usepackage{amsmath}            
\usepackage{comment}
\usepackage{latexsym}           
\usepackage{dcolumn}            
\usepackage[dvips]{graphicx}   
\usepackage{enumerate}
\usepackage{epstopdf}           
\usepackage{fancyhdr}           
\usepackage{hyperref}           
\usepackage{setspace}           
\usepackage{xspace}             
\usepackage{subfigure}          
\usepackage{bm}                 
\usepackage[ulem=normalem]{changes}

\newcommand{\bea}{\begin{eqnarray}}
\newcommand{\eea}{\end{eqnarray}}

\newcommand{\cf}{{\it cf.}~}

\newcommand{\eps}{\epsilon}
\newcommand{\etal}{{\it et al.}}
\newcommand{\ie}{{\it ie.\,}}
\newcommand{\bfx}{{\bf x}}
\newcommand{\taupd}{T_{pd}}
\newcommand{\al}{\alpha}

\begin{document}

\title{Shearing a glass and the role of pin-delay in models of interface depinning}
\date{\today}

\author{Stefanos Papanikolaou}
\affiliation{Departments of Mechanical Engineering \& Materials Science, Yale University, New Haven, Connecticut 06520}
\affiliation{Department of Physics, Yale University, New Haven, Connecticut 06520}

\begin{abstract}
When a disordered solid is sheared, yielding is followed by the onset of intermittent response that is characterized by slip in local regions usually labeled  shear-transformation zones (STZ). Such intermittent response resembles the behavior of earthquakes or contact depinning, where a well-defined landscape of pinning disorder prohibits the deformation of an elastic medium. Nevertheless, a disordered solid is evidently different in that pinning barriers of particles are due to  neighbors that are also subject to motion. Microscopic yielding leads to destruction of the local microstructure and local heating. It is natural to assume that locally a liquid emerges for a finite timescale before cooling down to a transformed configuration. For including this characteristic transient in glass depinning models, we propose a general mechanism that involves a ``pin-delay" time  $T_{pd}$, during which each region that slipped evolves as a fluid. The new timescale can be as small as a single avalanche time-step.
We demonstrate that the inclusion of this mechanism causes a drift of the critical exponents towards higher values for the slip sizes $\tau$, until a transition to permanent shear-banding behavior happens causing almost oscillatory, stick-slip response. Moreover, it leads to a proliferation of large events that are highly inhomogeneous and resemble sharp slip band formation. Our model appears to be qualitatively consistent with recent experiments and simulations of disordered solids under shear.
\end{abstract}

\maketitle 

\begin{figure}[t]
\includegraphics[width=0.46\textwidth]{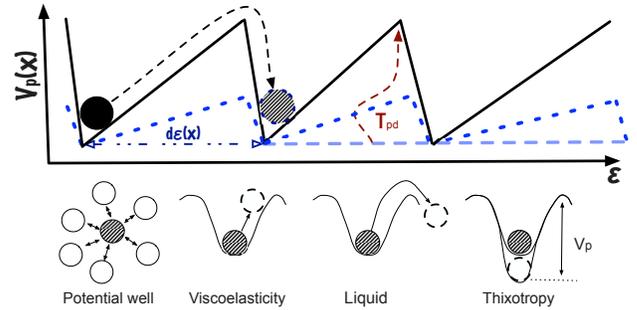}
\caption{
{\bf Thixotropic effects in depinning modeling}. Lower panel shows schematics how different mechanical states can be viewed in terms of their local potential well. Viscoelasticity brings particles back to their minima after the removal of the external stress, while in a liquid particles jump away from their minima. Thixotropy corresponds to the curious case where the minimum becomes deeper with time. Depinning is thus a liquid form; to add thixotropy, minima should become deeper after yielding.  Upper panel shows a typical random pinning potential for a single interface location $V_{p}(\bfx)$ as a function of plastic strain, as it is used in simulations. As external force increases, the interface slips by $d\eps(\bfx)$, the pinning potential disappears and then It reappears thixotropically (dashed lines) to its parent form during time $T_{pd}$. We consider, in simulations, the simplest case where at time  $T_{pd}$ the potential appears in a single step.
 }
 \label{fig:1}
\end{figure}

Extreme phenomena in nature appear in a wide range of scales, from the abrupt nano-deformation of materials to earthquake faults that extend several miles. At all scales, avoiding such phenomena requires a deep understanding of the disparate timescales leading to separate as well as collections of events. Material plastic deformation displays characteristic intermittency in a wide range of systems: single crystals~\cite{Dimiduk:2006uq}, bulk metallic glasses (BMG)~\cite{Sun:2010kx,Wang:2009vn,Shan:2008ys}, disordered granular solids~\cite{Richter:2010qf, Dahmen:2011bh}, colloids~\cite{Fall:2010dq}, frictional contacts~\cite{Burridge:1967cr, Fisher:1997nx}. While numerical simulations of materials can be very detailed at short timescales, the detailed study of a collection of abrupt events remains still out of reach. The theoretical intuition for such systems comes from the thorough study of {\it interface depinning}: a $d$-dimensional elastic string travelling on a landscape of quenched disorder in $d+1$ dimensions under the help of a uniform external force~\cite{Fisher:1998fk}. In complex disordered solids, though, while the assumption of an elastic medium appears reasonable, the one related to quenched disorder should be placed under scrutiny: the pinning disorder for every particle originates in the actual interface that attempts to depin (other nearby particles); a disordered solid pins itself during deformation. A basic consequence is that local glass depinning is not {\it immediately} followed by re-pinning; instead, in short-time transients (compared to typical avalanche durations), the system behaves locally as a fluid of finite viscosity $\eta$; In the context of depinning models we call this phenomenon {\it pin-delay}, generalizing the concept of complex fluid {\it thixotropy}~\cite{peterfi1927}: the change from a solid-like elastic gel to a flowable fluid as a function of time at fixed external stress.  
In this paper, we focus on the problem of plastic deformation in disordered solids and we investigate how pin-delay influences the statistics of avalanches and global strain behavior in a simple model.

Plastic deformation in disordered solids, such as BMGs or granular piles, has been a topic of interest for several decades~\cite{Schuh:2007ly, Volkert:2007,Hays:2000ve}; recently, it has become clear in well controlled experiments on BMGs that it does not evolve smoothly but through plastic bursts that organize into stick-slip stress oscillations with universal scaling~\cite{Richter:2010qf}. In a paradox, while BMGs are strongly disordered, plastic bursts along STZs spatially organize along sharp (nanometer-range) slip bands, a phenomenon coined {\it slip localization}. In simulations, avalanche statistics in both two and three dimensions~\cite{Salerno:2013kl} display universal behavior~\cite{Bailey:2007hc,Tewari:1999tg} with the critical exponent $\tau\simeq1.3$ if the particle ensemble follows steepest-descent dynamics. That universality class appears to be consistent with a simple, coarse-grained model of interface depinning~\cite{Talamali:2011ij,Vandembroucq:2011bs}. 
In retrospect, for less dissipative dynamics (a possible analog to increasing thixotropy) Salerno and Robbins~\cite{Salerno:2013kl} reported in both two and three dimensions a crossover where  the stress drops display different universal statistics with  $\tau\simeq1.6$ and larger and strongly anisotropic plastic bursts~\cite{Salerno:2012oq}.

Overdamped dynamics has been essential in the study of avalanche critical behavior, since numerical calculations remain tractable with event-driven dynamics being applicable. Departing away from that assumption to more complex local dynamics, such as underdamped finite inertia dynamics, is typically avoided. Finite inertia carries a system over successive (in deformation) pinning-potential energy minima to {\it minimize elastic interactions} and its effect appears similar to thixotropy in essence. The common approach used to include inertia effects is through emulating a frictional stick-slip mechanism, with rules that lower pinning barriers after the initial slip~\cite{Friedman:2012fk,Dahmen:2009uq,Prado:1992kx,Alava:2006vn,Dahmen:2011ys,Maimon:2004zr}, without successive minima jumps or direct elastic interaction minimization of any sort. All studies of this type find that addition of local  weakening introduces a global stick-slip instability discontinuously~\cite{Dahmen:2009uq,Prado:1992kx,Maimon:2004zr,Schwarz:2001ly}. Another approach to model inertia effects is through the inclusion of viscous effects in the elastic interaction, however the behavior is questionable near the depinning transition
\footnote{
We note that while there are attempts to model viscous effects~\cite{Marchetti:2000ve}, the approach used heavily relied on conventional calculus manipulations, even though near the depinning transition the dynamical equations are known to be intrinsically stochastic. Please see Refs.~\cite{LeDoussal12,Papanikolaou:2011fu} and references therein.
}. 
Finally, in mean-field approaches of interface depinning it is possible to study effective acceleration terms, but they do not alter the critical exponents and qualitative behavior near depinning~\cite{LeDoussal12}.

In this paper, we propose a mechanism where abrupt local slip leads to the disappearance of the local pinning potential for a characteristic timescale $T_{pd}$. This is a local thixotropic feature that we conjecture to be present in all disordered solids, independent of composition or scale of fluctuations.
During the interval $\taupd$, the system {\it flows} locally as a fluid of viscosity $\eta$ (considered a material property that can be thought as analog to Newtonian fluid viscosity, but defined for transients between slip events). 

We follow Talamali \etal's approach~\cite{Talamali:2011ij} for $d=2$ systems and assume that plastic deformation in disordered solids is modeled by the $xx-$component of the strain tensor $\eps\equiv\eps_{xx}=-\eps_{yy}$. The interaction due to local slip is the stress generated by local deformations of a random medium~\cite{Eshelby:1957fv}, which takes the form 
$\tilde F_{\rm int}(k,\omega)=(-\cos(4\omega)-1)\tilde \eps$ 
where $k,\;\omega$ are the polar coordinates in Fourier space~\cite{Budrikis:2013fk} and $\tilde\eps, \;\tilde F_{\rm int}$ are the transforms of the interaction and the strain. We scale the interaction strength by $c=0.1$, equivalent to modifying the strain-slip scale. We initialize the $L\times L$ system with $\eps(\bfx)=0\;\forall \;\bfx$ and stress thresholds $f_p(\bfx)$ taken from a uniform distribution $[0,1)$. We increase the external stress $F$ quasistatically until yielding, and then, the following evolution equation is solved,
\bea
\frac{d\eps(\bfx)}{dt}=\sigma(\bfx)=F_{\rm int}(\bfx) + F - f_p(\bfx) - k v(\bfx)
\eea
Assuming that $f_p(\bfx)$ corresponds to a pinning potential resembling the one in Fig.~\ref{fig:1} (upper panel), we implement a simple event-driven algorithm to integrate the above equation, in that there is a strain increase $d\eps(\bfx)$ randomly picked from a uniform $[0,1)$ distribution (characteristic of the pinning potential) when $\sigma(\bfx)>0$. The stress is decreased by $k v(\bfx)$ at each time-step, where $k$ is a local material weakening parameter and $v(\bfx)$ represents the fraction of STZs that yielded at the previous time-step.\footnote{This is not the only option for introducing a cut-off in the simulations~\cite{Papanikolaou:2011fu, Talamali:2011ij}, but it is appropriate for making contact with experimental results and simulations, given that we investigate the internal timescale structure of the model. Explicit comparison between models lies beyond the purpose of this work.} Thixotropic effects can be implemented in the following way: for every $\bfx$ that slipped at time $t$, we demand that $f_p(\bfx)=0$ for a pin-delay time-interval $T_{pd}$ that is an integer multiple of the unit time-step in an avalanche. 
Then, in the simplest possible ansatz, during the pin-delay time after the slip, the system locally evolves as a fluid, assuming a fluidity coefficient $D\equiv1/\eta$,
\bea
\frac{d\eps (\bfx)}{dt}= D\, \sigma(\bfx) 
\eea
When time $T_{pd}$ has lapsed or the avalanche stops, a random threshold force is picked. The avalanche stops when $\sigma(\bfx)<0 \;\forall\; \bfx$.

\begin{figure}[t]
\includegraphics[width=0.49\textwidth]{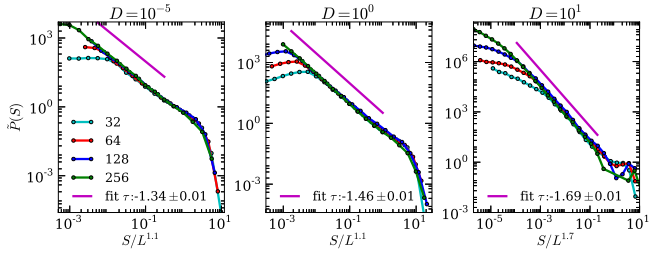}
\caption{
{\bf Scaling collapses of slip size probability distributions}. Size distributions are shown at $D=10^{-5},10^{0}$ and $10^1$ at $\taupd=8$, $k=0.2$. At high $D$, there is a clear drift of the critical exponent $\tau$ to $\sim1.7$, the cutoff scaling exponent also changes from $\sim 1.1$ to $\sim 1.7$ and the shape of the scaling function is clearly modified with an observable hump at large sizes. 
 }
\label{fig:2}
\end{figure}

\begin{figure}[t]
\includegraphics[width=0.49\textwidth]{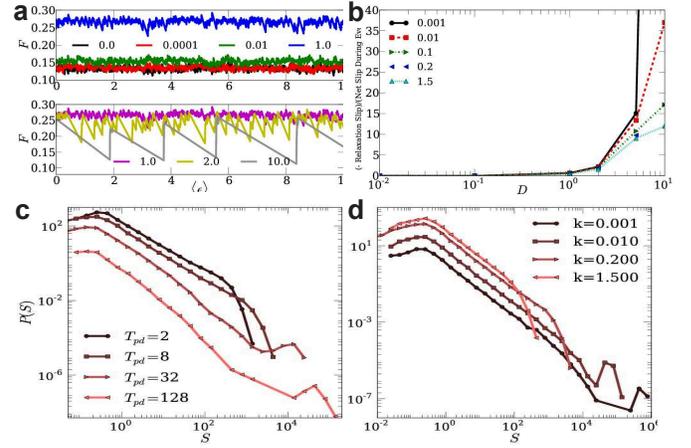}
\caption{
{\bf Effects of $D$, $T_{pd}$ and $k$ on serrated flow distributions}.
We set $L=128$. a. Evolution of serrated flow structure as $D$ increases. When $D<1$, its increase leads to a flow stress increase, due to competing effect of the relaxation. When $D>1$, stick-slip behavior is observed.
b. The fraction of slip during relaxation compared to net slip is shown. A transition is clear at $D\simeq1$, where the net slip is a clear outcome of the competition between relaxation and avalanches. Different plots correspond to different $k$.
c. The increase of $T_{pd}$ leads to exponent increase, and behavior that resembles the large $D$ behavior. Here $D=0.1$, $k=0.2$.
d. The increase of $k$ leads to reduction of the cutoff size.  Distributions are shifted by factors of $4$ for clarity.
}
\label{fig:3}
\end{figure}

We first investigate this model as a function of $D\equiv1/\eta$ for fixed $\taupd=8$. The external stress required to maintain the steady state behavior starts decreasing and develops stick-slip oscillations as $D>D^*\simeq1$~(\cf Fig.~\ref{fig:3}(upper left)). The critical exponent $\tau$ for the power law behavior of the slip size distribution increases from $\sim1.3$~\cite{Budrikis:2013fk} to $\sim1.52$ (and up to $1.65-1.7$ for $D>1$) and the form of the universal scaling function acquires a large-event hump (\cf Figs.~\ref{fig:2}, \ref{fig:3}). It is possible to perform a scaling collapse for the two different regimes at small and large $D$ (\cf Fig.~\ref{fig:2}): the behavior stays universal but the cutoff $S_0$ scales differently with the system size, $\sim L^{1.1}$ at low $D$ and $\sim L^{1.7}$ at high $D$. For intermediate $D\simeq1$, we observe a smooth exponent crossover. Avalanche durations (which are directly proportional to the stress drops in this model) display analogous behavior with their critical exponent changing from $1.6$ to $2.1$ and their cutoff exponent also changes from $\sim1.15$ to $\sim1.7$. For a scaling distribution $P(S)\sim S^{-\tau}$, that is trivially normalized $\int P(S,L)dS=1$, one has  $\tilde P(S)\equiv L^{\alpha}P(S)=g(S/L^{\alpha})$. Finally, we mention that at large $D$ the scaling collapse is not complete (the functional form at the cutoff region weakly changes) because an additional variable is present in the scaling function, $k$. The decrease of $k$ leads to a hump at larger sizes (\cf Fig.~\ref{fig:3} (right)) and with the current definitions, keeping $k$ fixed implies positive scaling of the average avalanche size with the system size~\cite{Talamali:2011ij}; thus, larger system sizes are closer to the parent depinning critical point; if we hold the distance from the critical point fixed (\ie by modifying $k$ to keep $\langle S\rangle$ fixed as the system size increases at $D=0$) then a multi-variable scaling collapse is possible, in the spirit of previous work on models of interface depinning~\cite{chen2011}.

It is interesting that the two different regimes compared to $D^*$ can be understood through an order parameter, in the amount of slip that took place during relaxation (Eq.~2) in comparison to the net slip (the sum of the result of Eqs.~1 and 2). The ratio of the relaxation slip over the net slip seems to display a strong increase from $\sim0$ at $D\simeq1$, and the increase becomes more drastic as $k$ is smaller and the system is closer to the critical point. (\cf Fig.~3b). The behavior of this ratio with $D$ seems to suggest a first-order transition scenario for the onset of the new $D>D^*$ regime.

For the behavior as a function of $\taupd$ for fixed $D$, we find that at low $D<D^*$, the increase of $\taupd$ leads to smaller avalanches  following the same parent distribution and similar critical exponents, acting in a sense like increasing $k$. This is sensible in that a large $\taupd$ does not allow avalanches to grow, cutting them off. However, the strength of $D$ is too small to build correlations among different STZs and thus, the critical exponents remain unaltered. However, as $D>D^*$ (\cf Fig.~\ref{fig:3}c) the increase of $\taupd$ leads to larger avalanches, higher exponent $\tau$ (up to $\sim1.7$) and a hump in the probability distribution that grows as $\sim 1/k$. 

When $k$ decreases, independently of the value of $D$, there appears to be a consistent drift of the critical exponent $\tau$ towards higher values, but the scaling function displays the characteristic hump only above $D^*$ (\cf Fig.~\ref{fig:3}d), and in that regime the critical exponents change drastically. The avalanche sizes at the hump scale as $\sim1/k$ and it appears plausible to conjecture that for any $D$ there is a small enough $k$ below which this novel behavior should be present, in accordance with the self-organized avalanche oscillator phenomenology~\cite{Papanikolaou:2012kl}.

\begin{figure}[t]
\includegraphics[width=0.49\textwidth]{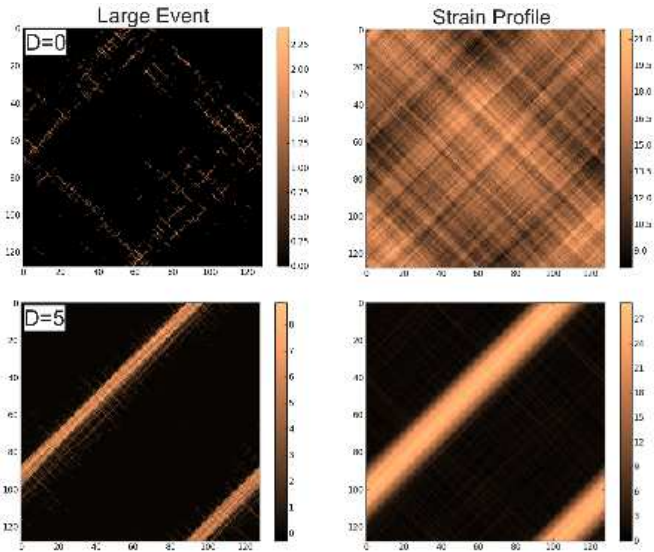}
\caption{{\bf Shear band formation due to pin-delay}. Here $L=128$, $k=0.2$, $T_{pd}=8$. The avalanche slip strain is shown on the left due to a typical large event near the cutoff of the distributions. The two dimensional strain-profile is shown on the right at the same strain as the one for the event shown.
{\bf Upper:} For small $D$, there are large events (left) but they appear non-spanning and not leading to shear-band formation~\cite{Talamali:2011ij}.
{\bf Lower:} Large events lead to shear-band formation in the strain profile.
 Large events at the hump location of the avalanche size distribution display sharp slip-band formation that is spanning along the diagonal, with thickness that decreases with system size. Our results are unaffected by a small diffusion term added to the elastic interaction for regularizing short-range interaction effects.}
\label{fig:4}
\end{figure}

Large events at the hump in the distributions at large $D$ display a non-trivial  system-spanning anisotropy along the diagonal, resembling sharp slip band formation (\cf Fig.~\ref{fig:4}(lower left)), in a strong amplification of the anisotropic features at $D=0$~\cite{Talamali:2011ij} (see also Fig.~\ref{fig:4}(upper left)). The features of Fig.~\ref{fig:4} (right) are qualitatively independent of short-range features of the interaction, since they persist in the presence of a small-amplitude laplacian/diffusion term in the interaction ($\sim k^2$). The pin-delay mechanism appears as a possible candidate to explain the onset of sharp shear bands in disordered solids under shear, as the strain profile is driven towards anisotropy as well (\cf Fig.~4(lower right)) in a consistent, discontinuous manner. It is worth noting that there are several suggestions for the onset of slip bands related to slow structural relaxations~\cite{Martens:2012zr,Jagla:2007ly}.

Experimentally, to our knowledge, there are no well accepted estimates for the critical exponents in avalanche behaviors for disordered solids, such as BMGs or colloidal systems. However, there are several promising experiments in the fields of granular/colloidal systems and BMGs that render support to our results. De Richter \etal~\cite{Richter:2010qf} in studies of sandpile avalanches at an incline observed a set of quasi-periodic bursts in addition to an apparent exponent drift for $\tau$, if compared to exponent values typically suggested~\cite{Dahmen:2011ys}. Further, Fall~\etal~\cite{Fall:2010dq} reported discontinuous slip band formation in colloidal systems with controlled increasing thixotropy. Also, the typical behavior of BMGs where slip band formation appears easier~\cite{Volkert:2007} at higher temperatures (where viscosity is lower) and smaller deformation rates (allowing for full relaxation during events), as naturally expected in our construction. While in support of our results, further experiments are needed to clarify the leading mechanisms for slip band formation and avalanches in disordered solids under shear, especially under controlled conditions such as colloidal systems with increasing thixotropy~\cite{Fall:2010dq} or variable interaction range~\cite{Guo:2011uq}. 

Our model's distinctive feature is that it contains time-intervals {\it during} the avalanche process where the pinning potential is locally absent. This mechanism resembles the avalanche oscillator mechanism~\cite{Papanikolaou:2012kl,Jagla:2010vn,Dobrinevski:2013ys} that functions during the long {\it waiting intervals} in-between events, in that there are time intervals where direct minimization of the stress is pursued, with no pinning involved. Nevertheless, the dynamics of our model is active only during avalanche events and is due to the disordered nature of the microstructure, while the avalanche oscillator is due to thermal relaxations (structural in nature) competing with avalanches to minimize elastic interactions. In comparison to our model, while frictional stick-slip mechanisms appear similar (both mechanisms assume that after local slip the pinning barrier drops), it leads to very different results, as we demonstrated. The differences are due to two key ingredients: 
i) the capacity of the system to locally seek the true stress minimum (aside from local barriers) for a small but finite timescale $\taupd$, 
ii) the fact that local slip during time $\taupd$ takes the system farther away from the local depinning threshold, albeit after having built collective, self-organized, correlations with the rest of the system.

\begin{acknowledgments}
I would like to thank M.~Robbins for bringing this problem to my attention and for very helpful and inspiring discussions. Also, I would like to thank S.~S.~Ashwin, K.~Dahmen, D.~M.~Dimiduk, J.~P.~Sethna, D.~Vandembroucq, E.~Van~der~Giessen and S.~Zapperi for inspiring discussions. This work is supported through a DTRA grant No. 1-10-0021 at Yale University. This work benefited greatly from the facilities and staff of the Yale University Faculty of Arts and Sciences High Performance Computing Center.
\end{acknowledgments}

\newpage
\section{Supplementary Material}
In this supplement, we discuss,
\begin{enumerate}
	\item How the model proposed can be written explicitly in terms of a depinning model equation of motion.
	\item In which ways the simulations of this paper are consistent with recent microscopic simulations.
	\item How the microscopic glass phenomenology is consistent and supports the phenomenology of this model.
	\item How a microscopically motivated model can be written based on the ideas of this work.
\end{enumerate}

\subsection{Thixotropy in depinning models}
It is possible to formally introduce the ``pin-delay" mechanism in depinning models by considering~(\cf\,Fig.~\ref{fig:1}) that the pinning term's strength $f_p$ is modified to become locally rate-dependent through $C_{pd}(t)=\frac{1}{\taupd}\int_{-\infty}^t dt' e^{-(t-t') / \taupd} ( 1-\Theta(\dot\eps(\bfx,t') - v_{th})  )$  or 
$C_{pd}(t)=\prod_{t'=t-\taupd}^{t}( 1-\Theta(\dot\eps(\bfx,t') - v_{th})  )$ (the latter applies directly to the model numerically solved in this paper) and the pinning term~\cite{Fisher:1998fk} is multiplied by $C_{pd}(t)$, while all the other stress terms are multiplied by $C_{pd}(t)+1/\eta(1-C_{pd}(t))$. When the system locally slips $\dot\eps>v_{th}$ (where $v_{th}$ is a threshold for viscoplastic flow rates),  it locally flows with viscosity $\eta$ for time $\taupd$ to minimize the local stress with no pinning force present. After the timescale $\taupd$ the system develops a usual pinning force, derived by a local pinning potential such as in Fig.~\ref{fig:1}. 
 
\subsection{Consistency with microscopic molecular dynamics simulations}
	Simulations in Refs.~\cite{Salerno:2012oq,Salerno:2013kl} showed that disordered solids undergoing underdamped dynamics enter a non-trivial universality class, where $\tau$ appears to show a crossover from $\sim1.3$ to $\sim1.6$ when it is estimated through the stress drops. This exponent crossover together with the qualitative features (hump) of the universal scaling function and the steady-state average stress consistent decrease, all resemble the behavior of our model. An apparent resemblance to the spatial features of the largest events is also observed. The exponent that characterizes the scaling of the distributions' cutoff $S_0$ with the system size (Salerno \etal\,label it $\alpha$) drifts from $\sim0.9$ to $\sim1.6$ in the underdamped regime; it resembles the drift observed here for the same exponent, from $\sim1.1$ to $\sim1.7$ (or for the stress drops/durations, from $\sim1.15$ to $\sim1.5$). We note that the actual values for the cutoff exponent depend drastically in the way used to cut-off avalanches. However, the relative drift of the exponent should be fully controlled by the same mechanism that controls the drift of $\tau$; the qualitative similarity of the drift renders further support to the ideas put forward in this work. 

Features of Salerno \etal's simulations that cannot be predicted by our model include: i) the fact that the energy drops appear to show the same critical exponent in both overdamped and underdamped limits and ii)  the energy/stress drop exponents apparently change their exponent near a crossover damping rate. First, we note that the energy drops is a feature that we cannot predict since the model we used does not have an explicit energy functional that is minimized. Moreover, the distinction between energy, strain and stress jumps is not transparent in our model. There are ways to study models that use energy functionals, making simulations more expensive~\cite{Papanikolaou:2012kl}, and this is a direction that shall be pursued.  Second, it is plausible that this feature is special to the particular approach used by Salerno \etal\, for the onset of pin-delay, since in the no-damping limit their model becomes Hamiltonian, a feature that might imply further behavioral complexity and additional dynamical symmetries. A more general, purely dissipative, approach for pin-delay  should be able to avoid additional complexity.

\subsection{Consistency with experimental phenomenology on bulk metallic glasses}
The model studied in this work includes two distinct forms of dynamics that are intertwined through the formation of discrete, abrupt events or avalanches. Experimentally, there is a lot to be said about relaxation properties of bulk metallic glasses, but it is generally well understood that, as in typical glassy systems, two types of relaxation modes exist: the $\al$-relaxation modes activated at the glass transition temperature and the $\beta$-relaxation modes activated at a somewhat lower than the glass transition temperature. While still unclear~\cite{beta2014}, there is a correlation of the former ($\al$) with large-particle, spatially correlated, slow motions, while the latter ($\beta$) associate with small-particle, fast and relatively uncorrelated slip. In the model of this paper, the STZ slip dynamics obviosuly corresponds to the $\beta$-relaxations while the slow, viscous relaxation after the fast slip can be thought to correspond to $\alpha$-like relaxations. The strong effect of the fluidity $D$ on the capacity of the system towards shear-banding flows resembles strongly the fragility parameter which signifies the dependence of the inverse relaxation time for the $\al$-relaxation modes on temperature ($m=d\ln(\tau_\al)/d\ln(T/T_g)$).

In addition, one can understand the analogy between existing models of glasses under shear by considering the basic phenomenology of granular systems under shear. For example, in the case of hard-disks under shear, the expectation is that the contact number locally displays a strong fluctuation with time, in between shear-events. While never explicitly observed, there is strong indication towards that direction[..].  

\subsection{Microscopically motivated models: The Little Free-Volume model}
Microscopically, what is missing from the physical picture of typical STZ models is ideas that are typical in free-volume models. For example, free volume theories~\cite{Heggen2005} suggest that there is free-volume increase at short times (dilation) during shear, while there is slow decrease at longer times. A typical model of the type suggests,
\bea
\frac{d\eps}{dt}\propto c_f\sigma
\eea
where $c_f$ is the density of the flow defects, while $\sigma$ is the stress, which in constitutive models it is assumed to be external, but for our purposes we assume that it is the stress felt by a single STZ locally. Following the experimental analysis of BMGs, it is clear that $c_f$ has a non-trivial evolution in time, where the onset of a strain-rate leads to a defect creation rate $dc_f/dt\sim c_f(\ln c_f)^2\dot\eps$ while then, the defect concentration tend to decay towards its ``equilibrium" value $dc_f/dt\propto-c_f(c_f-c_{eq})$. 

It is clear, that this physical picture aims to apply inside a single STZ of our model. Namely, one should consider that $f_p\propto 1-c_f\equiv$(concentration of non-flowing defects) in the large enough STZ. Then, $f_p$ has a creation rate $df_p/dt\propto c_f(c_f-c_{eq})$ and a destruction rate that takes place just after yielding $df_p/dt\propto-c_f(\ln c_f)^2\dot\eps$ with $c_f\propto 1-f_p$. This model has a very similar structure to the model solved in this paper, however it is more realistic and should be solved in a separate work, in comparison to existing experiments.

Such a model can be interpreted in terms of our construction in the following way: The  model of this paper is based on a simple version of the STZ modeling, where the effective temperature dynamics is intrinsically encoded in the dynamics of deformed regions. However, in the context of such phenomenology it is natural to expect that free-volume modeling applies to the length scale below the size of a single STZ, and consequently to timescales close to the time unit, exactly as we discussed.


\begin{thebibliography}{40}
\expandafter\ifx\csname natexlab\endcsname\relax\def\natexlab#1{#1}\fi
\expandafter\ifx\csname bibnamefont\endcsname\relax
  \def\bibnamefont#1{#1}\fi
\expandafter\ifx\csname bibfnamefont\endcsname\relax
  \def\bibfnamefont#1{#1}\fi
\expandafter\ifx\csname citenamefont\endcsname\relax
  \def\citenamefont#1{#1}\fi
\expandafter\ifx\csname url\endcsname\relax
  \def\url#1{\texttt{#1}}\fi
\expandafter\ifx\csname urlprefix\endcsname\relax\def\urlprefix{URL }\fi
\providecommand{\bibinfo}[2]{#2}
\providecommand{\eprint}[2][]{\url{#2}}

\bibitem[{\citenamefont{Dimiduk et~al.}(2006)\citenamefont{Dimiduk, Woodward,
  LeSar, and Uchic}}]{Dimiduk:2006uq}
\bibinfo{author}{\bibfnamefont{D.~M.} \bibnamefont{Dimiduk}},
  \bibinfo{author}{\bibfnamefont{C.}~\bibnamefont{Woodward}},
  \bibinfo{author}{\bibfnamefont{R.}~\bibnamefont{LeSar}}, \bibnamefont{and}
  \bibinfo{author}{\bibfnamefont{M.~D.} \bibnamefont{Uchic}},
  \bibinfo{journal}{Science} \textbf{\bibinfo{volume}{312}},
  \bibinfo{pages}{1188} (\bibinfo{year}{2006}).

\bibitem[{\citenamefont{Sun et~al.}(2010)\citenamefont{Sun, Yu, Jiao, Bai,
  Zhao, and Wang}}]{Sun:2010kx}
\bibinfo{author}{\bibfnamefont{B.~A.} \bibnamefont{Sun}},
  \bibinfo{author}{\bibfnamefont{H.~B.} \bibnamefont{Yu}},
  \bibinfo{author}{\bibfnamefont{W.}~\bibnamefont{Jiao}},
  \bibinfo{author}{\bibfnamefont{H.~Y.} \bibnamefont{Bai}},
  \bibinfo{author}{\bibfnamefont{D.~Q.} \bibnamefont{Zhao}}, \bibnamefont{and}
  \bibinfo{author}{\bibfnamefont{W.~H.} \bibnamefont{Wang}},
  \bibinfo{journal}{Physical Review Letters} \textbf{\bibinfo{volume}{105}},
  \bibinfo{pages}{035501} (\bibinfo{year}{2010}).

\bibitem[{\citenamefont{Wang et~al.}(2009)\citenamefont{Wang, Chan, Xia, Yu,
  Shen, and Wang}}]{Wang:2009vn}
\bibinfo{author}{\bibfnamefont{G.}~\bibnamefont{Wang}},
  \bibinfo{author}{\bibfnamefont{K.~C.} \bibnamefont{Chan}},
  \bibinfo{author}{\bibfnamefont{L.}~\bibnamefont{Xia}},
  \bibinfo{author}{\bibfnamefont{P.}~\bibnamefont{Yu}},
  \bibinfo{author}{\bibfnamefont{J.}~\bibnamefont{Shen}}, \bibnamefont{and}
  \bibinfo{author}{\bibfnamefont{W.~H.} \bibnamefont{Wang}},
  \bibinfo{journal}{Acta Materialia} \textbf{\bibinfo{volume}{57}},
  \bibinfo{pages}{6146} (\bibinfo{year}{2009}).

\bibitem[{\citenamefont{Shan et~al.}(2008)\citenamefont{Shan, Li, Cheng, Minor,
  Syed~Asif, Warren, and Ma}}]{Shan:2008ys}
\bibinfo{author}{\bibfnamefont{Z.~W.} \bibnamefont{Shan}},
  \bibinfo{author}{\bibfnamefont{J.}~\bibnamefont{Li}},
  \bibinfo{author}{\bibfnamefont{Y.~Q.} \bibnamefont{Cheng}},
  \bibinfo{author}{\bibfnamefont{A.~M.} \bibnamefont{Minor}},
  \bibinfo{author}{\bibfnamefont{S.~A.} \bibnamefont{Syed~Asif}},
  \bibinfo{author}{\bibfnamefont{O.~L.} \bibnamefont{Warren}},
  \bibnamefont{and} \bibinfo{author}{\bibfnamefont{E.}~\bibnamefont{Ma}},
  \bibinfo{journal}{Physical Review B} \textbf{\bibinfo{volume}{77}},
  \bibinfo{pages}{155419} (\bibinfo{year}{2008}).

\bibitem[{\citenamefont{de~Richter et~al.}(2010)\citenamefont{de~Richter,
  Zaitsev, Richard, Delannay, Ca{\"e}r, and Tournat}}]{Richter:2010qf}
\bibinfo{author}{\bibfnamefont{S.~K.} \bibnamefont{de~Richter}},
  \bibinfo{author}{\bibfnamefont{V.~Y.} \bibnamefont{Zaitsev}},
  \bibinfo{author}{\bibfnamefont{P.}~\bibnamefont{Richard}},
  \bibinfo{author}{\bibfnamefont{R.}~\bibnamefont{Delannay}},
  \bibinfo{author}{\bibfnamefont{G.~L.} \bibnamefont{Ca{\"e}r}},
  \bibnamefont{and} \bibinfo{author}{\bibfnamefont{V.}~\bibnamefont{Tournat}},
  \bibinfo{journal}{Journal of Statistical Mechanics: Theory and Experiment}
  \textbf{\bibinfo{volume}{2010}}, \bibinfo{pages}{P11023}
  (\bibinfo{year}{2010}).

\bibitem[{\citenamefont{Dahmen et~al.}(2011{\natexlab{a}})\citenamefont{Dahmen,
  Ben-Zion, and Uhl}}]{Dahmen:2011bh}
\bibinfo{author}{\bibfnamefont{K.~A.} \bibnamefont{Dahmen}},
  \bibinfo{author}{\bibfnamefont{Y.}~\bibnamefont{Ben-Zion}}, \bibnamefont{and}
  \bibinfo{author}{\bibfnamefont{J.~T.} \bibnamefont{Uhl}},
  \bibinfo{journal}{Nature Physics} \textbf{\bibinfo{volume}{7}},
  \bibinfo{pages}{554} (\bibinfo{year}{2011}{\natexlab{a}}).

\bibitem[{\citenamefont{Fall et~al.}(2010)\citenamefont{Fall, Paredes, and
  Bonn}}]{Fall:2010dq}
\bibinfo{author}{\bibfnamefont{A.}~\bibnamefont{Fall}},
  \bibinfo{author}{\bibfnamefont{J.}~\bibnamefont{Paredes}}, \bibnamefont{and}
  \bibinfo{author}{\bibfnamefont{D.}~\bibnamefont{Bonn}},
  \bibinfo{journal}{Physical Review Letters} \textbf{\bibinfo{volume}{105}},
  \bibinfo{pages}{225502} (\bibinfo{year}{2010}).

\bibitem[{\citenamefont{Burridge and Knopoff}(1967)}]{Burridge:1967cr}
\bibinfo{author}{\bibfnamefont{R.}~\bibnamefont{Burridge}} \bibnamefont{and}
  \bibinfo{author}{\bibfnamefont{L.}~\bibnamefont{Knopoff}},
  \bibinfo{journal}{Bulletin of the Seismological Society of America}
  \textbf{\bibinfo{volume}{57}}, \bibinfo{pages}{341} (\bibinfo{year}{1967}).

\bibitem[{\citenamefont{Fisher et~al.}(1997)\citenamefont{Fisher, Dahmen,
  Ramanathan, and Ben-Zion}}]{Fisher:1997nx}
\bibinfo{author}{\bibfnamefont{D.~S.} \bibnamefont{Fisher}},
  \bibinfo{author}{\bibfnamefont{K.}~\bibnamefont{Dahmen}},
  \bibinfo{author}{\bibfnamefont{S.}~\bibnamefont{Ramanathan}},
  \bibnamefont{and} \bibinfo{author}{\bibfnamefont{Y.}~\bibnamefont{Ben-Zion}},
  \bibinfo{journal}{Physical Review Letters} \textbf{\bibinfo{volume}{78}},
  \bibinfo{pages}{4885} (\bibinfo{year}{1997}).

\bibitem[{\citenamefont{Fisher}(1998)}]{Fisher:1998fk}
\bibinfo{author}{\bibfnamefont{D.~S.} \bibnamefont{Fisher}},
  \bibinfo{journal}{Physics Reports} \textbf{\bibinfo{volume}{301}},
  \bibinfo{pages}{113} (\bibinfo{year}{1998}).

\bibitem[{\citenamefont{Peterfi}(1927)}]{peterfi1927}
\bibinfo{author}{\bibfnamefont{T.}~\bibnamefont{Peterfi}},
  \bibinfo{journal}{Arch. Entwicklungsmech. Org.}
  \textbf{\bibinfo{volume}{112}}, \bibinfo{pages}{680} (\bibinfo{year}{1927}).

\bibitem[{\citenamefont{Schuh et~al.}(2007)\citenamefont{Schuh, Hufnagel, and
  Ramamurty}}]{Schuh:2007ly}
\bibinfo{author}{\bibfnamefont{C.~A.} \bibnamefont{Schuh}},
  \bibinfo{author}{\bibfnamefont{T.~C.} \bibnamefont{Hufnagel}},
  \bibnamefont{and}
  \bibinfo{author}{\bibfnamefont{U.}~\bibnamefont{Ramamurty}},
  \bibinfo{journal}{Acta Materialia} \textbf{\bibinfo{volume}{55}},
  \bibinfo{pages}{4067} (\bibinfo{year}{2007}).

\bibitem[{\citenamefont{Volkert et~al.}(2007)\citenamefont{Volkert, Cordero,
  Lilleodden, Donohue, and Spaepen}}]{Volkert:2007}
\bibinfo{author}{\bibfnamefont{C.}~\bibnamefont{Volkert}},
  \bibinfo{author}{\bibfnamefont{N.}~\bibnamefont{Cordero}},
  \bibinfo{author}{\bibfnamefont{E.}~\bibnamefont{Lilleodden}},
  \bibinfo{author}{\bibfnamefont{A.}~\bibnamefont{Donohue}}, \bibnamefont{and}
  \bibinfo{author}{\bibfnamefont{F.}~\bibnamefont{Spaepen}},
  \emph{\bibinfo{title}{Size {E}ffects in the {D}eformation of {M}aterials:
  {E}xperiments and {M}odeling}} (\bibinfo{publisher}{MRS Symposia Proceedings
  No 976E (Materials Research Society, Warrendale, PA, 2007)},
  \bibinfo{year}{2007}), pp. \bibinfo{pages}{0976--EE11--01}.

\bibitem[{\citenamefont{Hays et~al.}(2000)\citenamefont{Hays, Kim, and
  Johnson}}]{Hays:2000ve}
\bibinfo{author}{\bibfnamefont{C.}~\bibnamefont{Hays}},
  \bibinfo{author}{\bibfnamefont{C.}~\bibnamefont{Kim}}, \bibnamefont{and}
  \bibinfo{author}{\bibfnamefont{W.}~\bibnamefont{Johnson}},
  \bibinfo{journal}{Physical Review Letters} \textbf{\bibinfo{volume}{84}},
  \bibinfo{pages}{2901} (\bibinfo{year}{2000}).

\bibitem[{\citenamefont{Salerno and Robbins}(2013)}]{Salerno:2013kl}
\bibinfo{author}{\bibfnamefont{K.~M.} \bibnamefont{Salerno}} \bibnamefont{and}
  \bibinfo{author}{\bibfnamefont{M.~O.} \bibnamefont{Robbins}},
  \bibinfo{journal}{arXiv preprint arXiv:1309.1872}  (\bibinfo{year}{2013}).

\bibitem[{\citenamefont{Bailey et~al.}(2007)\citenamefont{Bailey, Schi{\o}tz,
  Lema{\^\i}tre, and Jacobsen}}]{Bailey:2007hc}
\bibinfo{author}{\bibfnamefont{N.~P.} \bibnamefont{Bailey}},
  \bibinfo{author}{\bibfnamefont{J.}~\bibnamefont{Schi{\o}tz}},
  \bibinfo{author}{\bibfnamefont{A.}~\bibnamefont{Lema{\^\i}tre}},
  \bibnamefont{and} \bibinfo{author}{\bibfnamefont{K.~W.}
  \bibnamefont{Jacobsen}}, \bibinfo{journal}{Physical Review Letters}
  \textbf{\bibinfo{volume}{98}}, \bibinfo{pages}{095501}
  (\bibinfo{year}{2007}).

\bibitem[{\citenamefont{Tewari et~al.}(1999)\citenamefont{Tewari, Schiemann,
  Durian, Knobler, Langer, and Liu}}]{Tewari:1999tg}
\bibinfo{author}{\bibfnamefont{S.}~\bibnamefont{Tewari}},
  \bibinfo{author}{\bibfnamefont{D.}~\bibnamefont{Schiemann}},
  \bibinfo{author}{\bibfnamefont{D.~J.} \bibnamefont{Durian}},
  \bibinfo{author}{\bibfnamefont{C.~M.} \bibnamefont{Knobler}},
  \bibinfo{author}{\bibfnamefont{S.~A.} \bibnamefont{Langer}},
  \bibnamefont{and} \bibinfo{author}{\bibfnamefont{A.~J.} \bibnamefont{Liu}},
  \bibinfo{journal}{Physical Review E} \textbf{\bibinfo{volume}{60}},
  \bibinfo{pages}{4385} (\bibinfo{year}{1999}).

\bibitem[{\citenamefont{Talamali et~al.}(2011)\citenamefont{Talamali,
  Pet{\"a}j{\"a}, Vandembroucq, and Roux}}]{Talamali:2011ij}
\bibinfo{author}{\bibfnamefont{M.}~\bibnamefont{Talamali}},
  \bibinfo{author}{\bibfnamefont{V.}~\bibnamefont{Pet{\"a}j{\"a}}},
  \bibinfo{author}{\bibfnamefont{D.}~\bibnamefont{Vandembroucq}},
  \bibnamefont{and} \bibinfo{author}{\bibfnamefont{S.}~\bibnamefont{Roux}},
  \bibinfo{journal}{Physical Review E} \textbf{\bibinfo{volume}{84}},
  \bibinfo{pages}{016115} (\bibinfo{year}{2011}).

\bibitem[{\citenamefont{Vandembroucq and Roux}(2011)}]{Vandembroucq:2011bs}
\bibinfo{author}{\bibfnamefont{D.}~\bibnamefont{Vandembroucq}}
  \bibnamefont{and} \bibinfo{author}{\bibfnamefont{S.}~\bibnamefont{Roux}},
  \bibinfo{journal}{Physical Review B} \textbf{\bibinfo{volume}{84}},
  \bibinfo{pages}{134210} (\bibinfo{year}{2011}).

\bibitem[{\citenamefont{Salerno et~al.}(2012)\citenamefont{Salerno, Maloney,
  and Robbins}}]{Salerno:2012oq}
\bibinfo{author}{\bibfnamefont{K.~M.} \bibnamefont{Salerno}},
  \bibinfo{author}{\bibfnamefont{C.~E.} \bibnamefont{Maloney}},
  \bibnamefont{and} \bibinfo{author}{\bibfnamefont{M.~O.}
  \bibnamefont{Robbins}}, \bibinfo{journal}{Physical Review Letters}
  \textbf{\bibinfo{volume}{109}}, \bibinfo{pages}{105703}
  (\bibinfo{year}{2012}).

\bibitem[{\citenamefont{Friedman et~al.}(2012)\citenamefont{Friedman, Jennings,
  Tsekenis, Kim, Tao, Uhl, Greer, and Dahmen}}]{Friedman:2012fk}
\bibinfo{author}{\bibfnamefont{N.}~\bibnamefont{Friedman}},
  \bibinfo{author}{\bibfnamefont{A.~T.} \bibnamefont{Jennings}},
  \bibinfo{author}{\bibfnamefont{G.}~\bibnamefont{Tsekenis}},
  \bibinfo{author}{\bibfnamefont{J.-Y.} \bibnamefont{Kim}},
  \bibinfo{author}{\bibfnamefont{M.}~\bibnamefont{Tao}},
  \bibinfo{author}{\bibfnamefont{J.~T.} \bibnamefont{Uhl}},
  \bibinfo{author}{\bibfnamefont{J.~R.} \bibnamefont{Greer}}, \bibnamefont{and}
  \bibinfo{author}{\bibfnamefont{K.~A.} \bibnamefont{Dahmen}},
  \bibinfo{journal}{Physical Review Letters} \textbf{\bibinfo{volume}{109}},
  \bibinfo{pages}{095507} (\bibinfo{year}{2012}).

\bibitem[{\citenamefont{Dahmen et~al.}(2009)\citenamefont{Dahmen, Ben-Zion, and
  Uhl}}]{Dahmen:2009uq}
\bibinfo{author}{\bibfnamefont{K.~A.} \bibnamefont{Dahmen}},
  \bibinfo{author}{\bibfnamefont{Y.}~\bibnamefont{Ben-Zion}}, \bibnamefont{and}
  \bibinfo{author}{\bibfnamefont{J.~T.} \bibnamefont{Uhl}},
  \bibinfo{journal}{Physical Review Letters} \textbf{\bibinfo{volume}{102}},
  \bibinfo{pages}{175501} (\bibinfo{year}{2009}).

\bibitem[{\citenamefont{Prado and Olami}(1992)}]{Prado:1992kx}
\bibinfo{author}{\bibfnamefont{C.~P.} \bibnamefont{Prado}} \bibnamefont{and}
  \bibinfo{author}{\bibfnamefont{Z.}~\bibnamefont{Olami}},
  \bibinfo{journal}{Physical Review A} \textbf{\bibinfo{volume}{45}},
  \bibinfo{pages}{665} (\bibinfo{year}{1992}).

\bibitem[{\citenamefont{Alava et~al.}(2006)\citenamefont{Alava, Nukala, and
  Zapperi}}]{Alava:2006vn}
\bibinfo{author}{\bibfnamefont{M.~J.} \bibnamefont{Alava}},
  \bibinfo{author}{\bibfnamefont{P.~K.} \bibnamefont{Nukala}},
  \bibnamefont{and} \bibinfo{author}{\bibfnamefont{S.}~\bibnamefont{Zapperi}},
  \bibinfo{journal}{Advances in Physics} \textbf{\bibinfo{volume}{55}},
  \bibinfo{pages}{349} (\bibinfo{year}{2006}).

\bibitem[{\citenamefont{Dahmen et~al.}(2011{\natexlab{b}})\citenamefont{Dahmen,
  Ben-Zion, and Uhl}}]{Dahmen:2011ys}
\bibinfo{author}{\bibfnamefont{K.~A.} \bibnamefont{Dahmen}},
  \bibinfo{author}{\bibfnamefont{Y.}~\bibnamefont{Ben-Zion}}, \bibnamefont{and}
  \bibinfo{author}{\bibfnamefont{J.~T.} \bibnamefont{Uhl}},
  \bibinfo{journal}{Nature Physics} \textbf{\bibinfo{volume}{7}},
  \bibinfo{pages}{554} (\bibinfo{year}{2011}{\natexlab{b}}).

\bibitem[{\citenamefont{Maimon and Schwarz}(2004)}]{Maimon:2004zr}
\bibinfo{author}{\bibfnamefont{R.}~\bibnamefont{Maimon}} \bibnamefont{and}
  \bibinfo{author}{\bibfnamefont{J.}~\bibnamefont{Schwarz}},
  \bibinfo{journal}{Physical Review Letters} \textbf{\bibinfo{volume}{92}},
  \bibinfo{pages}{255502} (\bibinfo{year}{2004}).

\bibitem[{\citenamefont{Schwarz and Fisher}(2001)}]{Schwarz:2001ly}
\bibinfo{author}{\bibfnamefont{J.}~\bibnamefont{Schwarz}} \bibnamefont{and}
  \bibinfo{author}{\bibfnamefont{D.~S.} \bibnamefont{Fisher}},
  \bibinfo{journal}{Physical Review Letters} \textbf{\bibinfo{volume}{87}},
  \bibinfo{pages}{096107} (\bibinfo{year}{2001}).

\bibitem[{\citenamefont{Le~Doussal et~al.}(2012)\citenamefont{Le~Doussal,
  Petkovi\ifmmode~\acute{c}\else \'{c}\fi{}, and Wiese}}]{LeDoussal12}
\bibinfo{author}{\bibfnamefont{P.}~\bibnamefont{Le~Doussal}},
  \bibinfo{author}{\bibfnamefont{A.}~\bibnamefont{Petkovi\ifmmode~\acute{c}\else
  \'{c}\fi{}}}, \bibnamefont{and} \bibinfo{author}{\bibfnamefont{K.~J.}
  \bibnamefont{Wiese}}, \bibinfo{journal}{Phys. Rev. E}
  \textbf{\bibinfo{volume}{85}}, \bibinfo{pages}{061116}
  (\bibinfo{year}{2012}).

\bibitem[{\citenamefont{Eshelby}(1957)}]{Eshelby:1957fv}
\bibinfo{author}{\bibfnamefont{J.~D.} \bibnamefont{Eshelby}},
  \bibinfo{journal}{Proceedings of the Royal Society of London. Series A.
  Mathematical and Physical Sciences} \textbf{\bibinfo{volume}{241}},
  \bibinfo{pages}{376} (\bibinfo{year}{1957}).

\bibitem[{\citenamefont{Budrikis and Zapperi}(2013)}]{Budrikis:2013fk}
\bibinfo{author}{\bibfnamefont{Z.}~\bibnamefont{Budrikis}} \bibnamefont{and}
  \bibinfo{author}{\bibfnamefont{S.}~\bibnamefont{Zapperi}},
  \bibinfo{journal}{arXiv preprint arXiv:1307.2135}  (\bibinfo{year}{2013}).

\bibitem[{pap()}]{papanikolaou13}
\bibinfo{howpublished}{S. Papanikolaou (in preparation)}.

\bibitem[{\citenamefont{Chen et~al.}(2011)\citenamefont{Chen, Papanikolaou,
  Sethna, Zapperi, and Durin}}]{chen2011}
\bibinfo{author}{\bibfnamefont{Y.-J.} \bibnamefont{Chen}},
  \bibinfo{author}{\bibfnamefont{S.}~\bibnamefont{Papanikolaou}},
  \bibinfo{author}{\bibfnamefont{J.~P.} \bibnamefont{Sethna}},
  \bibinfo{author}{\bibfnamefont{S.}~\bibnamefont{Zapperi}}, \bibnamefont{and}
  \bibinfo{author}{\bibfnamefont{G.}~\bibnamefont{Durin}},
  \bibinfo{journal}{Phys. Rev. E} \textbf{\bibinfo{volume}{84}},
  \bibinfo{pages}{061103} (\bibinfo{year}{2011}).

\bibitem[{\citenamefont{Papanikolaou et~al.}(2012)\citenamefont{Papanikolaou,
  Dimiduk, Choi, Sethna, Uchic, Woodward, and Zapperi}}]{Papanikolaou:2012kl}
\bibinfo{author}{\bibfnamefont{S.}~\bibnamefont{Papanikolaou}},
  \bibinfo{author}{\bibfnamefont{D.~M.} \bibnamefont{Dimiduk}},
  \bibinfo{author}{\bibfnamefont{W.}~\bibnamefont{Choi}},
  \bibinfo{author}{\bibfnamefont{J.~P.} \bibnamefont{Sethna}},
  \bibinfo{author}{\bibfnamefont{M.~D.} \bibnamefont{Uchic}},
  \bibinfo{author}{\bibfnamefont{C.~F.} \bibnamefont{Woodward}},
  \bibnamefont{and} \bibinfo{author}{\bibfnamefont{S.}~\bibnamefont{Zapperi}},
  \bibinfo{journal}{Nature} \textbf{\bibinfo{volume}{490}},
  \bibinfo{pages}{517} (\bibinfo{year}{2012}).

\bibitem[{\citenamefont{Martens et~al.}(2012)\citenamefont{Martens, Bocquet,
  and Barrat}}]{Martens:2012zr}
\bibinfo{author}{\bibfnamefont{K.}~\bibnamefont{Martens}},
  \bibinfo{author}{\bibfnamefont{L.}~\bibnamefont{Bocquet}}, \bibnamefont{and}
  \bibinfo{author}{\bibfnamefont{J.-L.} \bibnamefont{Barrat}},
  \bibinfo{journal}{Soft Matter} \textbf{\bibinfo{volume}{8}},
  \bibinfo{pages}{4197} (\bibinfo{year}{2012}).

\bibitem[{\citenamefont{Jagla}(2007)}]{Jagla:2007ly}
\bibinfo{author}{\bibfnamefont{E.~A.} \bibnamefont{Jagla}},
  \bibinfo{journal}{Physical Review E} \textbf{\bibinfo{volume}{76}},
  \bibinfo{pages}{046119} (\bibinfo{year}{2007}).

\bibitem[{\citenamefont{Guo et~al.}(2011)\citenamefont{Guo, Ramakrishnan,
  Harden, and Leheny}}]{Guo:2011uq}
\bibinfo{author}{\bibfnamefont{H.}~\bibnamefont{Guo}},
  \bibinfo{author}{\bibfnamefont{S.}~\bibnamefont{Ramakrishnan}},
  \bibinfo{author}{\bibfnamefont{J.~L.} \bibnamefont{Harden}},
  \bibnamefont{and} \bibinfo{author}{\bibfnamefont{R.~L.}
  \bibnamefont{Leheny}}, \bibinfo{journal}{The Journal of chemical physics}
  \textbf{\bibinfo{volume}{135}}, \bibinfo{pages}{154903}
  (\bibinfo{year}{2011}).

\bibitem[{\citenamefont{Jagla}(2010)}]{Jagla:2010vn}
\bibinfo{author}{\bibfnamefont{E.~A.} \bibnamefont{Jagla}},
  \bibinfo{journal}{Physical Review E} \textbf{\bibinfo{volume}{81}},
  \bibinfo{pages}{046117} (\bibinfo{year}{2010}).

\bibitem[{\citenamefont{Dobrinevski et~al.}(2013)\citenamefont{Dobrinevski,
  Le~Doussal, and Wiese}}]{Dobrinevski:2013ys}
\bibinfo{author}{\bibfnamefont{A.}~\bibnamefont{Dobrinevski}},
  \bibinfo{author}{\bibfnamefont{P.}~\bibnamefont{Le~Doussal}},
  \bibnamefont{and} \bibinfo{author}{\bibfnamefont{K.~J.} \bibnamefont{Wiese}},
  \bibinfo{journal}{Physical Review E} \textbf{\bibinfo{volume}{88}},
  \bibinfo{pages}{032106} (\bibinfo{year}{2013}).

\bibitem[{\citenamefont{Marchetti et~al.}(2000)\citenamefont{Marchetti,
  Middleton, and Prellberg}}]{Marchetti:2000ve}
\bibinfo{author}{\bibfnamefont{M.~C.} \bibnamefont{Marchetti}},
  \bibinfo{author}{\bibfnamefont{A.~A.} \bibnamefont{Middleton}},
  \bibnamefont{and}
  \bibinfo{author}{\bibfnamefont{T.}~\bibnamefont{Prellberg}},
  \bibinfo{journal}{Physical Review Letters} \textbf{\bibinfo{volume}{85}},
  \bibinfo{pages}{1104} (\bibinfo{year}{2000}).

\bibitem[{\citenamefont{Papanikolaou et~al.}(2011)\citenamefont{Papanikolaou,
  Bohn, Sommer, Durin, Zapperi, and Sethna}}]{Papanikolaou:2011fu}
\bibinfo{author}{\bibfnamefont{S.}~\bibnamefont{Papanikolaou}},
  \bibinfo{author}{\bibfnamefont{F.}~\bibnamefont{Bohn}},
  \bibinfo{author}{\bibfnamefont{R.~L.} \bibnamefont{Sommer}},
  \bibinfo{author}{\bibfnamefont{G.}~\bibnamefont{Durin}},
  \bibinfo{author}{\bibfnamefont{S.}~\bibnamefont{Zapperi}}, \bibnamefont{and}
  \bibinfo{author}{\bibfnamefont{J.~P.} \bibnamefont{Sethna}},
  \bibinfo{journal}{Nature Physics} \textbf{\bibinfo{volume}{7}},
  \bibinfo{pages}{316} (\bibinfo{year}{2011}).
  
\bibitem{beta2014} Hai Bin Yu, Wei Hua Wang, Hai Yang Bai, and Konrad Samwer, National Science Review {\bf 00},1 (2014).

\bibitem{Heggen2005} M. Heggen, F. Spaepen and M. Feuerbacher, Journal of Applied Physics 97, 033506 (2005).

\end{thebibliography}
\end{document}